\documentclass[aps,reprint,groupedaddress,longbibliography]{revtex4-1}
\usepackage{graphicx}
\graphicspath{{C:/Users/santo719/Desktop/PRL manuscript}}
\usepackage{amssymb}
\usepackage{hyperref}
\usepackage{color}
\hypersetup{
    colorlinks=true,
    linkcolor=blue,
    filecolor=blue,      
    urlcolor=blue,
    citecolor=blue,
}
\usepackage{gensymb}
\makeatletter
\def\NAT@def@citea{\def\@citea{\NAT@separator}}
\makeatother

\begin{document}

\title{Microscopic Investigation of Vortex Breakdown in a Dividing T-Junction Flow}


\author{San To Chan, Simon J. Haward, Amy Q. Shen}
\affiliation{Okinawa Institute of Science and Technology Graduate University, Onna, Okinawa 904-0495, Japan}


\date{\today}

\begin{abstract}
3D-printed microfluidic devices offer new ways to study fluid dynamics. We present the first clear visualization of vortex breakdown in a dividing T-junction flow. By individual control of the inflow and two outflows, we decouple the effects of swirl and rate of vorticity decay. We show that even slight outflow imbalances can greatly alter the structure of vortex breakdown, by creating a net pressure difference across the junction. Our results are summarized in a dimensionless phase diagram, which will guide the use of vortex breakdown in T-junctions to achieve specific flow manipulation.
\end{abstract}


\maketitle


\setlength{\parskip}{0em}
\fontdimen3\font=0.00001pt

\setlength{\parindent}{2ex} Microfluidic devices containing junctions serve as versatile platforms for studying many fundamental problems in physics, engineering and biology. These devices have been used to explore various flow phenomena, such as droplet generation~\cite{1}, liquid-liquid mixing~\cite{2}, and hydrodynamic trapping~\cite{25}. Moreover, they have been exploited for rheological measurements~\cite{3}, single-molecule studies~\cite{18,19}, and biophysical experiments~\cite{4,20}. Despite their simple geometries, junctions exhibit complex hydrodynamic behaviors such as the formation of vortices due to Dean instability~\cite{5,6,14,16} or spontaneous symmetry breaking~\cite{7}, particularly when inertial effects (as quantified by the Reynolds number \(Re\)) are dominant. Of particular interest is the dividing T-junction with square cross-section, for which the inflow splits in opposite directions through two symmetrical outlets. Here, axisymmetric vortex breakdown is predicted to occur when the inlet Reynolds number \(Re_{in}\) exceeds the critical threshold \(Re_c\)~\({\approx}\)~\(320\)~\cite{8,9,10,11}. \par

\setlength{\parindent}{2ex} Axisymmetric vortex breakdown, the development of bubble-like regions inside which the flow recirculates, is possible when vorticity decay is present in a swirling flow~\cite{21}. Early studies were conducted in cylindrical tubes~\cite{26,27,28,29,30,36,37,38,39} or by delta wings~\cite{41,42,43,44}, using dye-based flow visualization, Particle Tracking Velocimetry (PTV)~\cite{33} or Laser Doppler Anemometry (LDA)~\cite{34}. While much knowledge has been gained, there has still been no clear and unambiguous visualizations of the phenomenon, for it was difficult to capture the whole structure of vortex breakdown clearly in a single image using large experimental setups. For example, Bottausci and Petitjeans~\cite{45} used small jets of dye to visualize the external spiraling structure of vortex breakdown, but the relatively small internal structure was invisible due to dye diffusion. In principle, microfluidics allows microscopic investigation of different flow phenomena, which opens a new door to uncovering the full structure of vortex breakdown. However, to date vortex breakdown in microfluidic devices has only been shown via numerical simulations and inferred experimentally from qualitative observations of the trapping of particles and air bubbles~\cite{8,9,10,11}. A main aim of this work is to make the first-ever direct observation of the vortex breakdown, and to provide new insight into its physical mechanism.\par

\begin{figure}[b]
\includegraphics[width=8.5cm]{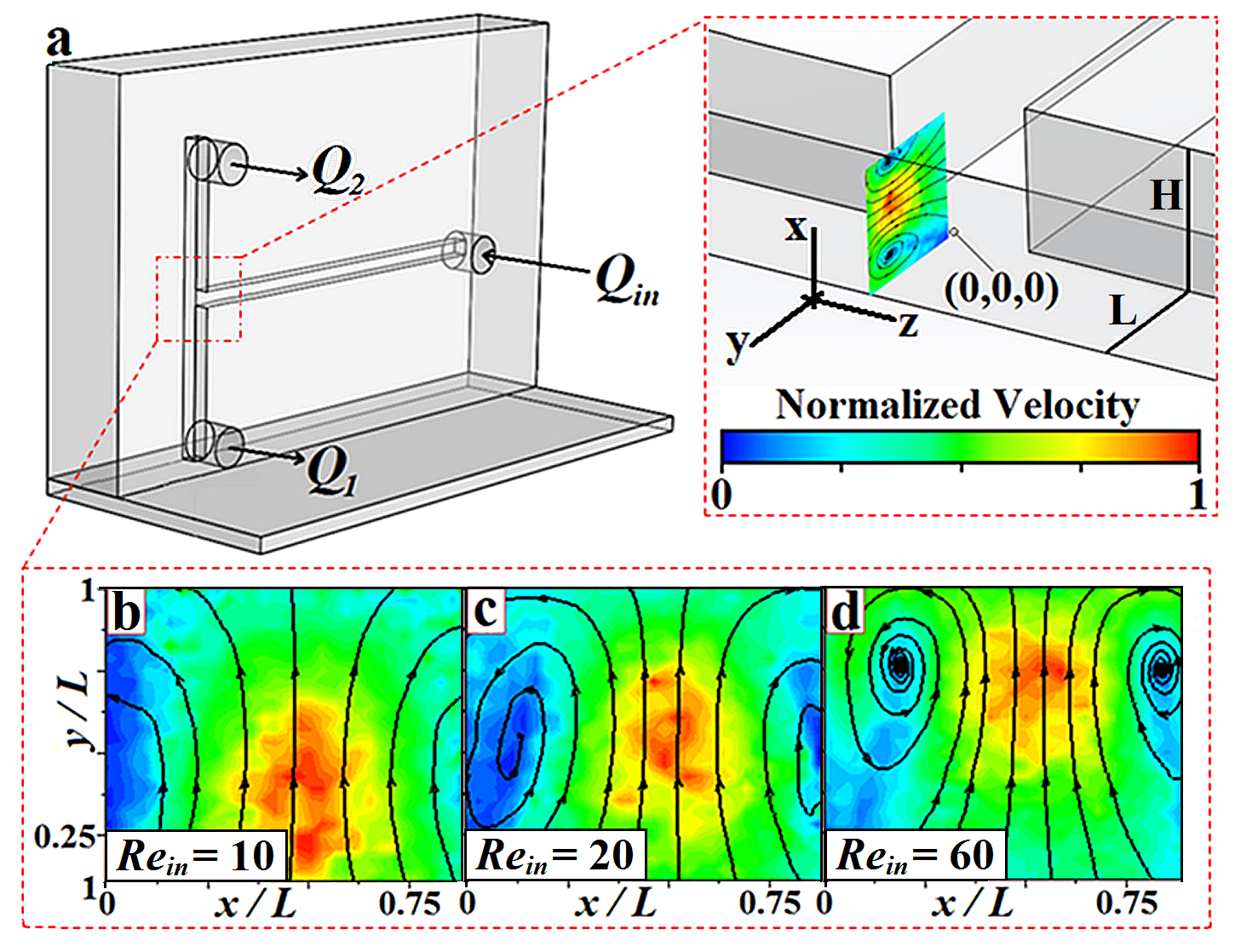}%
\centering
\caption{Schematic of the glass microfluidic T-junction device.~(a)~The inflow is denoted by \(Q_{in}\), the outflows towards outlets 1 and 2 are denoted by \(Q_1\) and \(Q_2\). The inlet Reynolds number is defined as \(Re_{in}\)~\({=}\)~\(\rho_fQ_{in}/\mu L\).~[(b)-(d)]~Micro-Particle Image Velocimetry~(\(\mu\)PIV) on the \(x\)-\(y\) plane showing vortex formation as \(Re_{in}\) is increased.\label{f1}}
\end{figure}

\setlength{\parindent}{2ex} We use a unique setup that allows multiple planes of observation and high-contrast imaging of recirculating streamlines in a microfluidic device, to report the full structure of vortex breakdown in a dividing T-junction flow. We effectively control the vorticity decay in the two outlets for a fixed vorticity at the junction center, by manipulating the two fundamental flow parameters that govern the vortex breakdown in T-junctions: the inflow and the ratio of the two outflows. This is important for many flow systems, since outflow imbalances can easily arise due to imprecise control of the flowrates, or asymmetries in the channel dimensions and pressure drops. First, we show by micro-Particle Image Velocimetry (\(\mu\)PIV)~\cite{13} how vortices form at the junction, and that the vorticity at the junction center can be treated as a constant when the ratio of the two outflows is varied. Such variation of even a few percent can alter the structures of the vortex breakdown and the critical Reynolds number. Finally, we perform numerical simulations to gain further insights into the mechanism of the vortex breakdown. \par

\setlength{\parindent}{2ex} The microfluidic T-junction device [Fig.~\hyperref[f1]{1(a)}], with square cross-section of sides \(L\)~\({=}\)~\(H\)~\({=}\)~\(420\)~\({\pm}\)~\(10\)~\(\mu\)m was fabricated in glass by Selective Laser-induced Etching~(SLE)~\cite{12}. The device was mounted vertically on a glass slide to image the flow through the \(x\)-\(y\) or \(y\)-\(z\) plane. The inlet length was \(25L\), which ensures the flow is fully developed before the junction. The outlet lengths were \(10L\), corresponding to the working distance of the microscopes. Combining this device with \(\mu\)PIV and flow visualization provides access to the formation process and breakdown structure of the vortices. We denote the inflow as \(Q_{in}\), and the respective outflows towards outlets~1 and~2 as \(Q_1\) and \(Q_2\). We use the inlet Reynolds number \(Re_{in}\)~\({=}\)~\(\rho_fQ_{in}/\mu L\), where \(\rho_f\) is the fluid density and \(\mu\) is the viscosity, to describe the inlet flow velocity in a dimensionless form. Similarly, we use \(Re_{1}\)~\({=}\)~\(\rho_fQ_{1}/\mu L\) and \(Re_{2}\)~\({=}\)~\(\rho_fQ_{2}/\mu L\) for the two outflows. We define an imbalance index \(I\)~\({=}\)~\((Re_1\)~\({-}\)~\(Re_2)/Re_{in}\) to quantify the level of imbalance between the two outflows. When \(I\)~\({=}\)~\(0\), the outflows are balanced. When \(I\)~\({=}\)~\(1\), the inflow is directed towards outlet 1 completely. To control \(Re_{in}\) and \(I\) simultaneously, the T-junction flow was driven by three individually controlled neMESYS syringe pumps~(Cetoni~GmbH,~Germany), which are equipped with \(10\)~ml glass syringes (Hamilton~Gastight,~Reno,~NV). One of the syringes varies \(Q_{in}\) to control \(Re_{in}\), while the other two vary \(Q_1\) and \(Q_2\) to control \(I\). The pumps were operated at a minimum of \(15{\times}\) the specified lowest pulsation-free flowrate, above which the syringe pistons are moved without vibration.\par

\setlength{\parindent}{2ex} For \(\mu\)PIV, red fluorescent polystyrene particles of diameter \(d_p\)~\({=}\)~\(2\)~\(\mu\)m and density \(\rho_p\)~\({=}\)~\(1.05\)~g/cm\(^3\) were seeded in deionized water and volumetrically illuminated by a \(527\)~nm dual-pulsed Nd:YLF~laser~(Terra~PIV,~Continuum~Inc.,~CA) with averaged power of \(60\)~W, repetition rate of \(500\)~Hz and pulse duration of \(10\)~ns. The time separation between laser pulses was \(1\)~\(\mu\)s~\({<}\)~\(\delta t\)~\({<}\)~\(400\)~\(\mu\)s depending on the flowrate. Images were captured by a high speed CMOS camera~(Phantom~Miro~M310,~Vision~Research~Inc.,~NJ) connected to an inverted microscope~(Nikon~ECLIPSE~T\(i\)-S) with a \(0.3\)~NA~\(10{\times}\)~objective. The measurement depth~\cite{22} over which particles contribute to the determination of the flow field was \(\delta z_m\)~\({\approx}\)~\(35\)~\(\mu\)m~\({<}\)~\(0.1L\). \par

\setlength{\parindent}{2ex}For flow visualization to trace flow streamlines, green fluorescent polystyrene particles of diameter \(d_p\)~\({=}\)~\(2\)~\(\mu\)m and density \(\rho_p\)~\({=}\)~\(1.05\)~g/cm\(^3\) were seeded in \(0.2\)~M sodium metatungstate solution of density \(\rho_f\)~\({=}\)~\(1.44\)~g/cm\(^3\) and shear viscosity \(\mu\)~\({=}\)~\(0.0013\)~Pa\(\cdot\)s. Particles were volumetrically illuminated by a metal halide lamp with a \(488\)~nm excitation filter, their trajectories were recorded with exposure time of \(20\)~ms by a spinning-disk confocal imaging system~(DSD2,~Andor Technology Ltd) connected to an inverted microscope~(Nikon~ECLIPSE~T\(i\)). For a wider field of view, a \(0.13\)~NA~\(4{\times}\)~objective was used, the spatial resolution was \(2.7\)~\(\mu\)m/pixel. While for a zoomed-in view, a \(0.3\)~NA~\(10{\times}\)~objective was used, the spatial resolution was \(1\)~\(\mu\)m/pixel. 
The density ratio \(\rho_p/\rho_f\)~\({\approx}\)~\(0.7\) ensures that the number of trapped particles is low \cite{8}. As the particles have Stokes number \(St\)~\({=}\)~\(\rho_pd_p^2Re_{in}/18\rho_fL^2\)~\({<}\)~\(0.001\) over the range of \(Re_{in}\) we considered, they trace the streamlines with negligible error~\cite{23}. By continuity, streamlines from the inlet must connect to the outlets, which ensures that the particles stay in the vortex breakdown regions temporarily. Also, as the flow has lower velocity in the vortex breakdown regions, more fluorescent signals from the particles can be collected locally per frame. The above factors enhance the contrast of the particle trajectories in the vortex breakdown regions, which permits us to visualize the recirculating streamlines clearly. Tracing the vortex breakdown streamlines would otherwise be impossible by using air bubbles, as they accumulate easily due to their large size and low density~\cite{8}, which disturbs the flow field. \par

\setlength{\parindent}{2ex} To gain deeper insight, we also performed finite volume, steady state, single phase flow simulations using ANSYS Fluent. The inlet and outlets of our T-junction model have square cross-section of side \(L\)~\({=}\)~\(1\) and channels of length \(5L\). The model was divided into half due to symmetry over the \(x\)~\({=}\)~\(L/2\) plane, and discretized into \(9900000\) elements. To correctly capture boundary layer effects, we refined the elements near the channel walls. The flow profile was fully developed before the junction. Further increasing the outlet lengths or resolution of mesh results in no significant difference in the flow patterns. \par

\setlength{\parindent}{2ex} Before discussing vortex breakdown, it is helpful to first understand why vortices form within a T-junction. We performed \(\mu\)PIV on the \(z\)~\({=}\)~\(0\) plane with \(I\)~\({=}\)~\(0\) (i.e., balanced outflows) [Figs.~\hyperref[f1]{1(b)}-\hyperref[f1]{1(d)}]. Increasing \(Re_{in}\) shifts the peak of the velocity distribution to the channel wall located at \(y\)~\({=}\)~\(L\) and creates a pair of counter-rotating vortices by the Dean mechanism \cite{14}. As the flow changes direction at the junction, centrifugal force is generated, which drives the fluid and the peak of the velocity distribution to the channel wall. This induces high pressure at the wall and the induced pressure gradient then drives the fluid backward, thus forming the observed pair of vortices. \par

\setlength{\parindent}{2ex} To investigate how outflow imbalances affect the Dean vortices, we measured the maximum vorticity of the \(\mu\)PIV flow field, on \(z\)~\({=}\)~\(0.1L\), for \(I\)~\({=}\)~\((0, \pm0.1, \pm0.2\)) and fixed \(Re_{in}\)~\({=}\)~\(500\) [Fig.~\hyperref[f1]{2(a)}]. The maximum vorticity corresponds to the vorticity of the vortex core. Measuring the vorticity for \(I\) on \(z\)~\({=}\)~\(0.1L\) is equivalent to measuring the vorticity for \(-I\) on \(z\)~\({=}\)~\(-0.1L\), due to the mirror symmetry over \(z\)~\({=}\)~\(0\). The data points of \(I\)~\({=}\)~\((\pm0.1, \pm0.2\)) enclose that of \(I\)~\({=}\)~\(0\). By the intermediate value theorem, within a narrow distance of \(0.2L\), there exists at least one point on the vortex core of \(I\)~\({=}\)~\((\pm0.1, \pm0.2\)) for which the vorticity is equal to that of \(I\)~\({=}\)~\(0\). In other words, we can interpret that the vorticity is fixed at the junction center when \(I\) is varied. To further justify this statement, we probed the core vorticities along the z-direction for \(I\)~\({=}\)~\(0\) and \(I\)~\({=}\)~\(0.2\) by both \(\mu\)PIV and simulation [Fig.~\hyperref[f1]{2(b)}]. Both results show that there exists a point between \(z\)~\({=}\)~\(-0.1L\) and \(z\)~\({=}\)~\(0.1L\) (orange circle in Fig.~\hyperref[f1]{2(b)}) where the core vorticities for \(I\)~\({=}\)~\(0\) and \(I\)~\({=}\)~\(0.2\) are equal. Hence, we can interpret that the effect of varying \(I\) is to change the rate of vorticity decay in the two outlets at a fixed vorticity at the junction center. This provides a simple way to study how vorticity decay affects the vortex breakdown in a T-junction flow. \par

\begin{figure}[b]
\includegraphics[width=8.5cm]{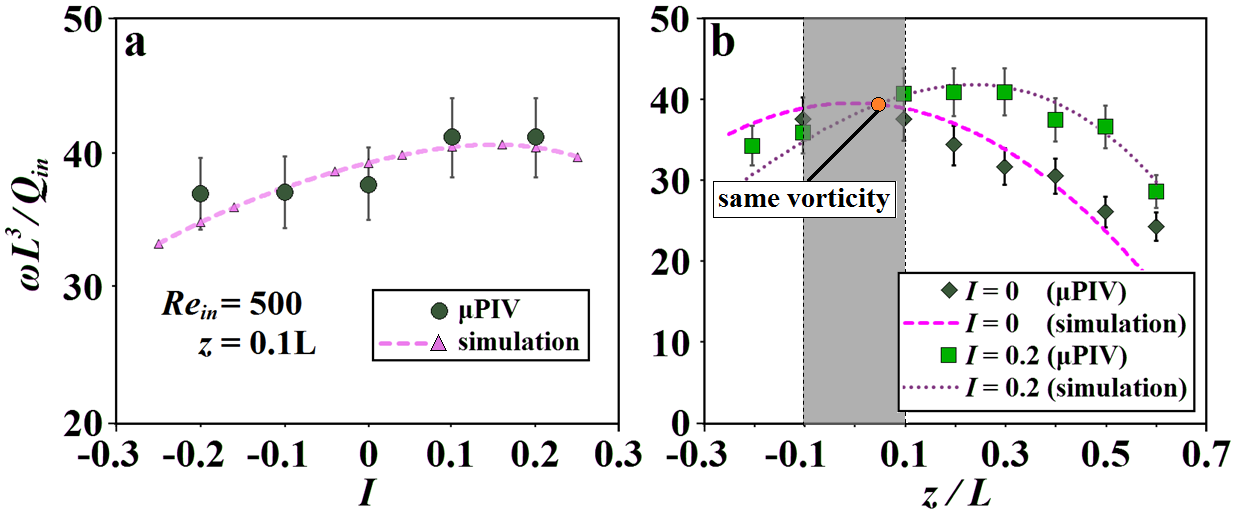}%
\centering
\caption{Vorticity along the vortex core probed by \(\mu\)PIV and simulation at \(Re_{in}\)~\({=}\)~\(500\).~(a)~Vorticity as a function of outflow imbalance \(I\) on the \(z\)~\({=}\)~\(0.1L\) plane. The data point of \(I\)~\({=}\)~\(0\) is enclosed by others. As a consequence of the intermediate value theorem, one can interpret that the vorticity at the junction center is fixed when \(I\) is varied.~(b)~Vorticity as a function of \(z\) for \(I\)~\({=}\)~\(0\) and \(I\)~\({=}\)~\(0.2\). In a distance of \(0.2L\) (gray area), these exists a point (orange circle) where the core vorticities for \(I\)~\({=}\)~\(0\) and \(I\)~\({=}\)~\(0.2\) are equal. \label{f2}}
\end{figure}

\begin{figure}[b]
\includegraphics[width=8.5cm]{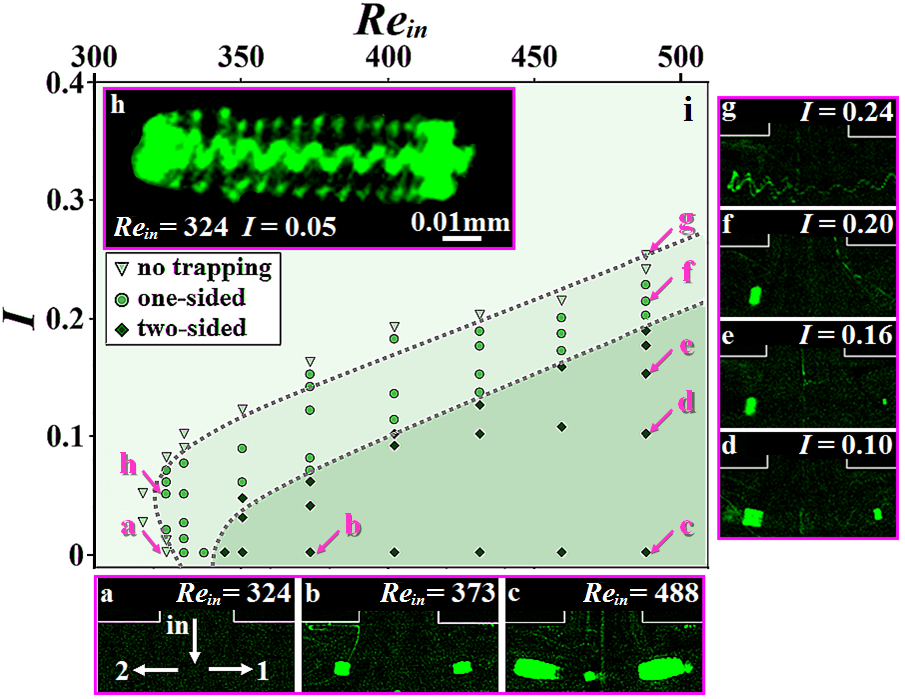}%
\centering
\caption{Particle trapping in a T-junction flow.[(a)-(c)]~\(I\)~\({=}\)~\(0\), \(Re\) increases: two symmetrical trapping regions emerge at the two outlets; their sizes increase with \(Re_{in}\).~[(d)-(f)]~\(Re_{in}\)~\({=}\)~\(465\), \(I\) increases: the trapping region in outlet~1 progressively shrinks until it disappears.~(g)~\(I\) increases further: the trapping region in outlet~2 disappears; particles retract and travel to outlet~1.~(h)~A particle's trajectory signifying vortex breakdown.~(i)~Phase diagram of particle trapping. \label{f3}}
\end{figure}

\setlength{\parindent}{2ex} Next, to study how outflow imbalances affect vortex breakdown, we visualized the particle trajectories on the \(y\)-\(z\) plane (Fig.~\hyperref[f2]{3}). With fixed \(I\)~\({=}\)~\(0\) and \(Re_{in}\)~\({>}\)~\(336\), we observed two symmetrical particle trapping regions in the two outlets [Figs.~\hyperref[f2]{3(a)}-\hyperref[f2]{3(c)}], whose size increases with \(Re\). Here, the onset \(Re_{in}\) for particle trapping is consistent with the values reported previously~\cite{8,10}, where the T-junctions span over several length scales by combining our results. However, our results differ from those of Ault~\textit{et}~\textit{al.}~\cite{10}, for their observed trappings were mainly asymmetric, due to the accumulation of trapped air bubbles, which perturbs the flow field. With fixed \(Re_{in}\)~\({=}\)~\(488\) and increasing \(I\), the trapping region in outlet~1 progressively shrinks until it vanishes at \(I\)~\({=}\)~\(0.2\) [Figs.~\hyperref[f3]{3(d)}-\hyperref[f3]{3(f)}]. Increasing \(I\) further, the trapping region in outlet~2 also vanishes, the particles retract and then travel to outlet~1 [Fig.~\hyperref[f3]{3(g)}]. Remarkably, we find trapping can be induced even for \(Re_{in}\)~\({<}\)~\(336\) by increasing \(I\) (see Fig.~\hyperref[f3]{3(h)} for \(I\)~\({=}\)~\(0.05\)). Zooming into the trapping region [Fig.~\hyperref[f3]{3(h)}], we clearly observe that the particle trajectory resembles the vortex breakdown structure predicted numerically by Chen~\textit{et}~\textit{al.}~\cite{11}. Vigolo~\textit{et}~\textit{al.}~\cite{8} also tracked the trajectories of air bubbles in the the T-junction, which showed spiraling shapes. However, our visualization is the first to simultaneously show both the internal and external spiraling structures of vortex breakdown. A movie corresponding to Fig.~\hyperref[f3]{3(a)-3(h)} can be found in the Supplemental Material~\cite{35}. \par

\setlength{\parindent}{2ex} To summarize our experimental results, we constructed a phase diagram in the (\(Re_{in}\)-\(I\)) plane based on the observed number of particle trapping regions (Fig.~\hyperref[f3]{3(i)}). Here, our data extends the phase diagram in the \(Re_{in}\)-\(\theta\) plane presented by Ault~\textit{et}~\textit{al.}~\cite{10} into a new dimension at junction angle \(\theta\)~\({=}\)~\(90\degree\). Increasing \(Re_{in}\) extends the range of \(I\) for which particles can get trapped. For \(Re_{in}\)~\({>}\)~\(336\) and different \(I\), the trapping can be either one-sided or two-sided. As \(I\) is increased from \(0\), two-sided trapping always occurs before one-sided trapping. For \(Re_{in}\)~\({<}\)~\(336\), one-sided trapping still occurs within a narrow range of \(I\). As the particle trapping phenomenon is described by two dimensionless parameters \(Re_{in}\) and \(I\), our phase diagram is applicable to all T-junctions having a square cross-section. Additionally, we have constructed two phase diagrams for T-junctions with the same inlet cross-sectional aspect ratio \(\alpha_{in}\)~\({=}\)~\(L_{in}/H\)~\({=}\)~\(1\) but outlet cross-sectional aspect ratios \(\alpha_{out}\)~\({=}\)~\(L_{out}/H\)~\({=}\)~\(0.5\) and \(2\) (Figs. S1-S2). They are qualitatively similar to Fig.~\hyperref[f3]{3}, there exists a range of \(Re_{in}\) and \(I\) for which one-sided trapping occurs. However, we notice that \(\alpha_{out}\) modifies the flow separation at the junction corner, which in turns affects the orientation of the particle trapping regions. This constitutes a new physical phenomenon highlighting the wall-liquid interaction that deserves further investigation. The two additional phase diagrams are included in the Supplemental Material~\cite{35}. \par

\begin{figure}[h]
\includegraphics[width=8.5cm]{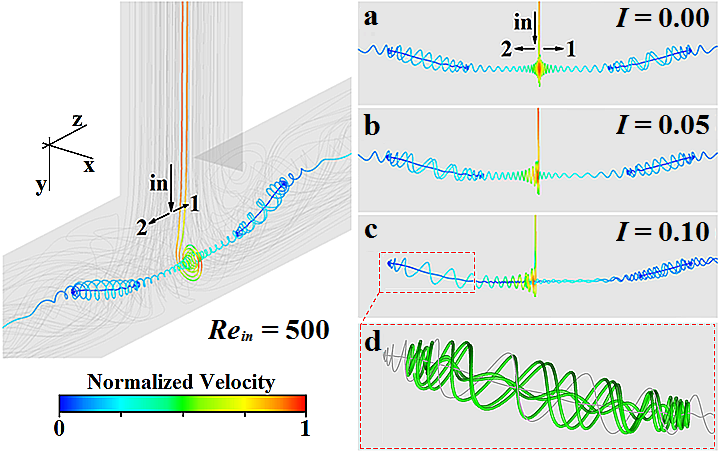}%
\centering
\caption{Steady state flow simulations of the T-junction flow with \(Re_{in}\)~\({=}\)~\(500\). [(a),~(b)]~Increasing \(I\) shrinks and enlarges the recirculation zones in outlet~1 and~2, respectively.~(c)~Further increasing \(I\), the vortex core of the recirculation zone in outlet~2 merges with the flow directing towards outlet~1.~(d)~The streamlines surrounding the merged vortex core in outlet~2 are still recirculating. \label{f4}}
\end{figure}

\setlength{\parindent}{2ex} With good agreement between the experimentally measured and numerically predicted vorticities [Fig.~\hyperref[f2]{2(a)-2(b)}], we now use the T-junction model to relate our particle trapping observations to the vortex breakdown in the two outlets. The simulation result for \(Re_{in}\)~\({=}\)~\(500\) and \(I\)~\({=}\)~\(0\) [Fig.~\hyperref[f4]{4(a)}] agrees with that of Chen~\textit{et}~\textit{al.}~\cite{11}, for the flow recirculation zones in the two outlets are symmetric and bubble-like, which signifies axisymmetric vortex breakdown. Increasing \(I\), the recirculation zone in outlet~1 shrinks [Fig.~\hyperref[f4]{4(b)}], corresponding to the particle trapping results [Figs.~\hyperref[f3]{3(d)}-\hyperref[f3]{3(f)}]. By contrast, the recirculation zone in outlet~2 enlarges, until its vortex core crosses \(z\)~\({\approx}\)~\(0\) and merges with the flow directing towards outlet~1 [Fig.~\hyperref[f4]{4(c)}]. We note that the streamlines surrounding the merged vortex core are recirculating [Fig.~\hyperref[f4]{4(d)}], which means particles can still be trapped. As the recirculation zone in outlet~2 keeps enlarging, more recirculating streamlines connect to outlet~1, which generates the particle retraction behavior [Fig.~\hyperref[f3]{3(g)}]. \par

\begin{figure}[h]
\includegraphics[width=8.5cm]{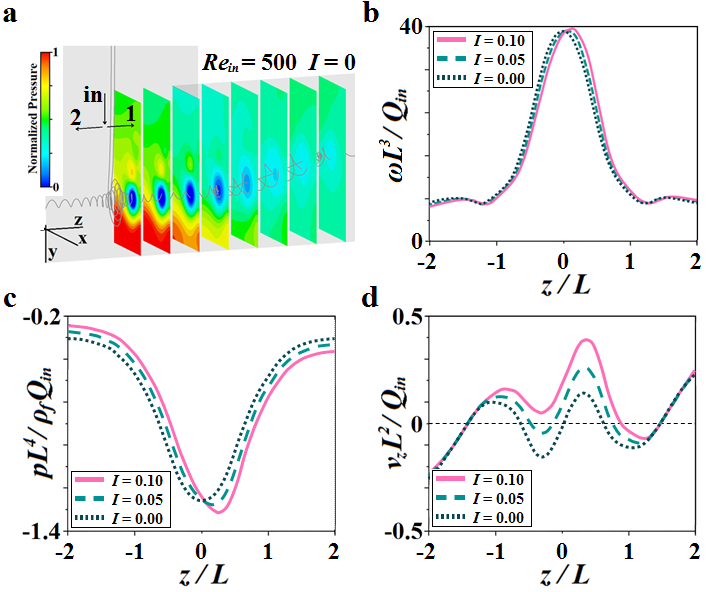}%
\centering
\caption{Numerically predicted pressure and velocity distributions with fixed \(Re_{in}\)~\({=}\)~\(500\).~(a)~Pressure contours along the \(z\)-direction, showing the decay of radial pressure gradient downstream due to viscous dissipation, which in turn creates an adverse pressure gradient.~(b)~Distribution of vorticity along the vortex core.~(c)~Distribution of pressure along the vortex core. ~(d)~Distribution of the \(z\)-component of the flow velocity along the vortex core. \label{f5}}
\end{figure}

\setlength{\parindent}{2ex} Essentially all the observations above can be explained by the change in pressure profiles in the two outlets. As the inflow impacts the channel wall, a pair of vortices is generated. Centrifugal force is induced, creating a radial pressure gradient along the vortex core. Due to viscous dissipation, the radial pressure gradient decays downstream, which in turn creates an adverse axial pressure gradient [Fig.~\hyperref[f5]{5(a)}]. Together with the contribution from flow deceleration due to flowrate halving, this adverse pressure gradient can be strong enough to cause local flow recirculation (i.e., vortex breakdown). Here the effect of increasing \(I\) is to decrease and increase the flow deceleration in outlets~1 and~2, respectively. This shifts the vorticity distribution [Fig.~\hyperref[f5]{5(b)}] and creates a net pressure difference between the two outlets [Fig.~\hyperref[f5]{5(c)}], the effect of which is to distort the velocity distribution along the vortex core [Fig.~\hyperref[f5]{5(d)}]. This changes the number and positions of the internal stagnation points and thus the number and structures of the recirculation zones. \par

\setlength{\parindent}{2ex} In a dividing T-junction flow, the deceleration between the inflow and outflows is constrained by the channel geometry, and the vorticity is coupled to the inflow by the Dean mechanism. Simply by varying the outflow imbalance, we decoupled all these factors to gain a fundamental understanding of the vortex breakdown. Our unique use of a 3D-printed microfluidic device to observe the flow from multiple perspectives can be applied to other fluid dynamics studies. Further, our phase diagram will guide designing devices to manipulate flow conditions using vortex breakdown. Finally, as inertial microfluidics is becoming more important in manipulating fluids in lab-on-a-chip devices~\cite{15}, our data shows the importance of accurate channel design and flow control to avoid failure in microfluidic experiments involving flow deceleration and vorticity decay. \par

\end{document}